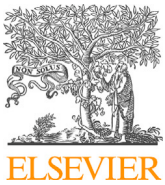

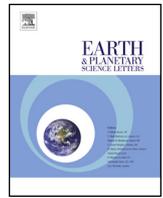

# Chondrites as thermal and mechanical archives of accretion processes in the Solar protoplanetary disk

Anthony Seret [a], Guy Libourel [a,b,*]

[a] *Laboratoire Lagrange, Observatoire de la Côte d'Azur, boulevard de l'Observatoire, Nice, CS 34229 - F 06304 Nice Cedex 4, France*
[b] *Hawai'i Institute of Geophysics and Planetology, School of Ocean, Earth Science and Technology, University of Hawai'i at Mānoa, Honolulu, 96821, Hawai'i, USA*

## ARTICLE INFO

## ABSTRACT

As some of the most ancient materials in our Solar System, chondritic meteorites offer a valuable window into the early stages of planetary formation, particularly the accretion processes that built the most primitive asteroids. Until now, high energy shocks and collisions have been invoked to explain the deformation and fragmentation of chondrules, the main component of chondrites. However, simulating the cooling of chondrules using continuum mechanics and finite elements, we demonstrate that plastic deformation of chondrules can occur at low collision velocities of just a few meters per second and with kinetic energies less than tenths of a millijoule when temperatures exceed the glass transition temperature $T_g \sim 1000\,\mathrm{K}$. Conversely, below $T_g$, spontaneous chondrule cracking occurs due to differential thermal contraction between phases and is more pronounced in larger chondrules. Counterintuitively, our findings suggest that both ordinary and carbonaceous chondrites formed through similar low-energy processes, with varying degrees of ductility and brittleness depending on the amount of processed material. This implies that the environments where chondrites formed were likely less turbulent and more thermally active than previously thought.

## Introduction

Chondrites are the most primitive meteorites formed during the first five million years of the Solar System evolution (Krot et al., 2005; Villeneuve et al., 2009; Connelly et al., 2012; Doyle et al., 2015) and their major element compositions come close to the composition of the Sun (Lodders, 2003). Hosting in their fine-grained matrix, presolar grains, refractory inclusions and chondrules as the oldest solids of the Solar System as well as various types of organic matter, chondrites have come under increasing scrutiny (Scott and Krot, 2014). Bulk compositions, and physical and textural features of chondrites result from accretion of their parent asteroids and their subsequent lithification from metamorphism and/or aqueous alteration due to internal heating by the decay of $^{26}Al$ short-lived radionuclide and shock heating associated with impacts. Despite providing unique information on a whole array of astrophysical and cosmochemical processes, chondrites remain still enigmatic cosmic sediments, whose accretion is not well understood.

Most chondrites have been classified into distinct groups based on similarities in petrological, chemical and isotopic characteristics (Krot et al., 2005). Of the several groups of ordinary and carbonaceous chondrites, all but carbonaceous-Ivuna type (CI) group contain chondrules (Fig. 1). Chondrules, the main component of chondrites, are believed to have formed by melting at high temperatures (1600 K to 2100 K) of solid precursors in dust-rich zones of the protoplanetary disk, followed by cooling during which they crystallized and solidified to form rigid silicate droplets (Scott and Krot, 2014). This paper focuses on type 3 chondrites, which have undergone minimal thermal and aqueous alteration. Chondrules are classified into three textural types: porphyritic, nonporphyritic, and glassy (Jones et al., 2018). Ordinary and carbonaceous chondrites differ by several of their petrological characteristics, including chondrule abundance, size as well as amount of matrix (Krot et al., 2005). Ordinary chondrites are characterized by abundant chondrules (up to 80-90 % vol.) and some chondrule/mineral fragments dispersed in an < 10 % vol. inter-chondrule fine-grained matrix. H-L-LL ordinary chondrites host cluster chondrite clasts that include close-fit textures of deformed and indented chondrules (Metzler, 2012; Yamaguchi et al., 2019; Goudy, 2019; Ruzicka et al., 2015, 2020). Previous investigations revealed that about 40 % of unequilibrated ordinary chondrites (UOCs) contain cluster chondrite clasts with modal abundances between 5 % vol. and 90 % vol. The original dimension of the rocks parental to these clasts is unknown; the largest clast found so far has a diameter of 10 cm, confined by the size of the meteorite (Metzler, 2012). In these chondrule aggregates, a significant fraction of chondrules deviates from an ideal spherical shape as expected for solidification of molten droplets in low gravity conditions and appear to be deformed in a viscous state by indentation of neighboring chondrules (Fig. 1a) (Metzler, 2012). In contrast, CR, CO, CM and CV carbonaceous chondrites are characterized by various amounts of undeformed chondrules (20 % to 50 %), som of which being lobate (Jacquet, 2021), numerous fragments of different types of chondrules and minerals such as isolated pyroxene and olivine (Fig. 1b and supplementary figure S6), scattered in an abundant fine-grained interchondrule matrix (30 % to 70 % vol.) (Krot et al., 2005). Ordinary and carbonaceous chondrites are the two main classes of chondrites, which are thought to originate from two genetically distinct reservoirs in the inner and outer solar system, respectively, (Warren, 2011; Kruijer et al., 2020).

Different modes of formation have been put forward for explaining the textural differences between chondrites (Fig. 1). For cluster chondrites, this includes accretion of hot and viscous chondrules, jostling and wedging during chondrite compaction, short reheating by thermal metamorphism, or uniaxial deformation during shock on their respective parent body (Metzler, 2012). Fragments of chondrules, on the other hand, are interpreted as resulting from collisional destruction due to localized turbulence in the solar nebula or high-velocity impacts on meteorite parent bodies (Metzler, 2012; Nelson and Rubin, 2010; Ueda et al., 2001; Arakawa and Nakamoto, 2016). Considering from the above that formation of chondrites comes down to understanding how chondrules can deform or break into pieces, we followed a mechanistic approach based on continuum mechanics as the overarching theory and finite elements simulations for practical resolution (see Methods). We show how the mechanical behavior of chondrules during cooling differs on both sides of the glass transition temperature ($T_g$), which helps to explain the various textures observed in ordinary and carbonaceous chondrites and put tight constraints on chondrite accretion mechanism.

## 1. Methods

Chondrules as partially crystallized silicate spherules are inferred to form from molten dust balls by rapid cooling because they commonly contain glass with variable composition (Hewins et al., 2005; Jones et al., 2018). As pointed out for magma solidification, dynamic cooling induces complex interactions between crystallization processes and rheological behavior, which are key to understand how chondrules behave prior to their accretion. A key physical quantity is here the glass transition temperature defined as the temperature above which the material, although solid, presents a ductile mechanical behavior much like a liquid, and below which the materials presents a brittle mechanical behavior as a solid uncapable of ductile deformation. Accordingly, our simulations using continuum mechanics and finite elements (see supplementary material) model the mechanical behavior of partially crystallized chondrules above and below the glass transition temperature of the mesostasis circa $T_g = 1000$ K, assuming that the glass transition takes place when the glass specific heat approaches the Dulong-Petit limit around $3 \cdot R$, where $R$ is the universal gas constant (Richet et al., 2021). Finite elements simulations have been carried out using the Abaqus software (Dassault Systems) notably able to model nonlinear cases due to variable material behavior laws and/or contacts as well as static load cases.

Leveraging Abaqus's versatility, our current investigation aims to identify the specific stress and deformation conditions required to replicate deformed chondrules in ordinary chondrites and fragmented chondrules in carbonaceous chondrites. To achieve this, we have simplified the chondrule model by representing the chondrule mesostasis as a sphere containing a variable-sized, spherical olivine inclusion positioned either at the center or at the edge. The modeled chondrule composition is limited to Mg-rich composition (mimicking olivine-rich Type I chon-drules), specifically pure forsterite for the olivine and a representative type I chondrule mesostasis for the surrounding glassy shell ((Libourel et al., 2006) and supplementary material). Collisions between chondrules at temperatures exceeding their glass transition temperature $T_g$ were simulated using aggregates with volumes ranging from low to typical values observed in cluster chondrite clasts of ordinary chondrites (around 1 cm³). The simulations incorporated collision velocities up to 40 m s⁻¹, reflecting the assumption that maximum velocities caused by radial drift and turbulence within the protoplanetary disk typically range from 8 m s⁻¹ to 100 m s⁻¹ (Arakawa et al., 2023). Spontaneous cracking inside chondrules below $T_g$ have been simulated by considering modeling the propagation of a spherical cap shaped crack along the mesostasis-forsterite interface. Different chondrule sizes and forsterite positions have been tested to study the likelihood of chondrules to crack. Mechanical and thermal properties of forsterite and chondrule glassy mesostasis used as input parameters are detailed in the supplementary material.

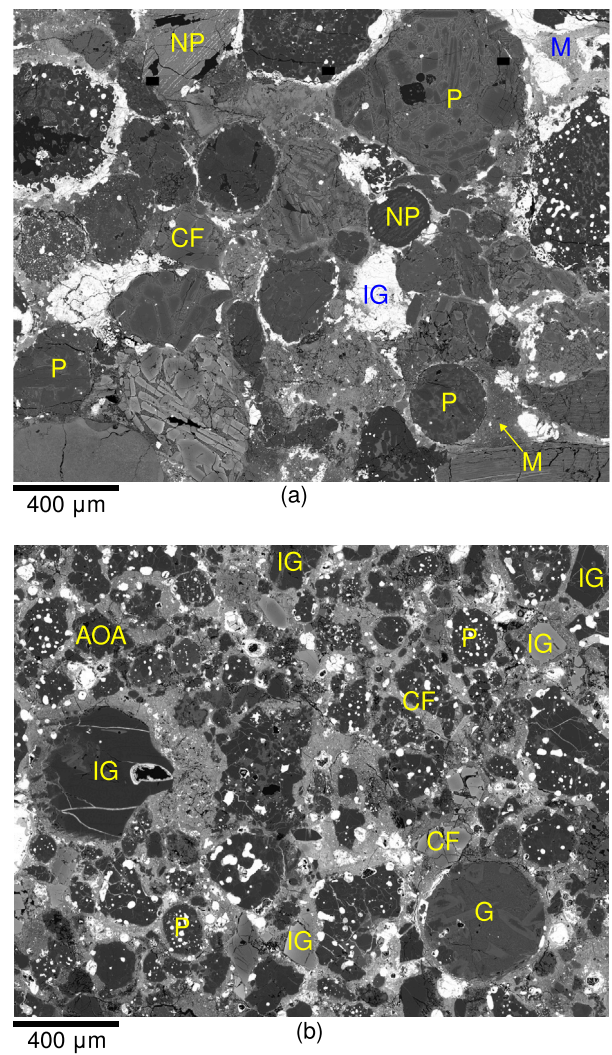

**Fig. 1.** Backscattered electron images from scanning electron microscope of (a) Semarkona LL3.0 ordinary chondrite and (b) Yamato 81020 CO 3.05 carbonaceous-Ornans type chondrite. Ordinary and carbonaceous chondrites differ by several of their petrological characteristics, including types of chondrules (porphyritic (P), Non-porphyritic (NP) and glassy (G)), their abundance and size. They also differ by their texture (see text). Notice the close-fit textures of deformed and indented chondrules and the relative low proportion of interchondrule matrix (M) in Semarkona as opposed to the large abundance of fragments of chondrules (CF) and isolated grains (IG) in Yamato 81020 and the higher proportion of matrix. Isolated grains are mainly composed of fractured olivine grains of different compositions, some of which have some glassy mesostasis attached to them.

## 2. Results

### 2.1. Chondrule mechanical behavior above the glass transition temperature, $T_g \sim 1000\,\text{K}$

In ordinary chondrites, deformed chondrules are observed modal abundances of 40 % chondrules having experienced a deformation greater than 20 % i.e. NWA 1756 LL3.1 chondrite (Metzler, 2012). These results have been interpreted as evidence of accretion by collisions at high temperature where the mesostasis of chondrules behaves like a liquid above its glass transition temperature, and thus capable of ductile deformation.

These conditions have been simulated by describing the rheological behavior of a partially crystallized chondrule as part of an aggregate of rigid and glued chondrules upon impact with an aggregate where all chondrules are identical in size and positions but are all rigid and glued together (Fig. 2). The parameters are a temperature of 1373 K (Holmén and Wood, 1986) and relative velocity of $40\,\text{m}\,\text{s}^{-1}$ (Arakawa et al., 2023).

The main difference in this simulation is the mechanical property of the mesostases of both colliding chondrules: liquid with compressibility (isotropic behavior linking pressure and density change through the bulk modulus, see supplementary material) and temperature-dependent viscosity (deviatoric behavior) against perfectly rigid (top and bottom part respectively in the scheme of colliding aggregates in Fig. 2). The simulations were carried out at an increasing number (1, 4, 16, 64, 256) of chondrules per aggregate. There is no noticeable deformation of the mesostasis of chondrules, even for aggregates made of 64 chondrules, when the kinetic energy involved is insufficient. However, with an aggregate of 256 chondrules (corresponding to a mass of 37 mg and a kinetic energy of 7.4 mJ) the chondrule deformation is characterized by an indented mesostasis shape similar to those observed in cluster chondrites (Figs. 1 and 2c).

Collisions under the same conditions have also been simulated with each aggregate containing $2^{14} = 16384$ chondrules resulting in a total volume of approximately $1\,\text{cm}^3$, which is typical of observed clasts with deformed chondrules (Metzler, 2012). By systematically increasing the relative velocity starting from 1 meter/s, no significant deformation of the non-rigid chondrule mesostasis until reaching 8 meter/s (kinetic energy of 38 mJ for the combined aggregates) has been observed. At this critical velocity, chondrule deformation manifested as an indented mesostasis morphology (Fig. 3), resembling those observed in cluster chondrites (Figs. 1 and 2c).
With the ability to explore a wider range of aggregate velocity, mass, and temperature combinations, these simulations could provide not only qualitative visualizations but also quantitative analyses of the resulting chondrule morphology. Furthermore, the requirement for a large number of chondrules per aggregate aligns perfectly with observations of numerous chondrules within ordinary chondrite cluster chondrites (Metzler, 2012). This strong concordance suggests that these simulations effectively capture the core mechanisms driving the deformation process.

Both the amount of strain (Figs. 2b and 3) and the temperature are heterogeneous in the mesostasis, and increase near deformation areas (olivine-mesostatis interface and inter-mesostases contact), with temperature increasing by 200 K (collision of $1\,\text{cm}^3$ aggregates at $8\,\text{m}\,\text{s}^{-1}$ relative velocity, rightmost case in Fig. 3). However, maximal values are not the most reliable metric as they result from local concentrated effects which may not be properly quantified if the spatial and/or temporal mesh is not extremely fine.

### 2.2. Chondrule mechanical behavior below the glass transition temperature, $T_g \sim 1000\,\text{K}$

Possible cracking inside chondrules below $T_g$ has been simulated by describing the propagation of an initial crack, i.e. a Griffith flaw (Griffith, 1920) chosen here as a 1 μm-long crack along the interface between an elastic sphere of forsterite and an elastic mesostasis spherical shell (Fig. 4(a, b)). Different chondrules sizes and forsterite position have been tested. The effect of chondrule size on cracking was studied by considering five chondrule sizes: a standard size with a mesostasis diameter of 500 μm and a forsterite diameter of 100 μm, and four other sizes obtained by homotheties of factors 2, 4, 8, and 16, producing chondrules with diameters of 1 mm, 2 mm, 4 mm and 8 mm, respectively. The larger chondrules mimic macro-chondrules. We also tested the influence of forsterite position in the mesostasis for a standard-sized chondrule by considering a forsterite placed at the border of the mesostasis.

Results of crack propagation upon cooling are illustrated qualitatively by magnifying the crack front at two different positions throughout propagation for the standard-sized chondrule (Fig. 4(c, d)). The angular position i.e. colatitude $\theta$ (defined as shown by orange lines on Fig. 4b where it is equal to $\theta_{ini}$ for the initial crack) of the crack front is plotted as a function of the temperature upon cooling for the central (supplementary figure S2a) and border forsterite (supplementary figure S2b) for the standard chondrule size to show the effect of the forsterite position in the mesostasis. It is shown that chondrules with central olivine grains are more susceptible to crack propagation than those with edge olivine grains. This is because the greater amount of mesostasis surrounding central olivine grains allows for higher storage and release of elastic strain energy, which is the driver of crack propagation.

Results of crack propagation are quantitatively illustrated in Fig. 5 by plotting the angular position i.e. colatitude $\theta$ of the crack front as a function of temperature and chondrule size for the central forsterite. In particular, crack propagation is easier as chondrule size increases because the driving force for crack propagation, elastic strain energy, scales with the cube of chondrule size, while the hindering force, crack surface energy, scales with the square of the forsterite size. As a result, elastic strain energy becomes more important relatively to interface energy as chondrule size increases, which promotes crack propagation. Moreover, the crack propagates almost entirely throughout the forsterite-mesostasis interface within a temperature interval of at most 100 K upon cooling (represented by horizontal rectangular bars with gradient of color in Fig. 5) which supports the fact that such initial crack (or more generally, defect) is instable and suddenly propagates almost entirely to fracture.

## 3. Discussion

As a matter of fact, chondrules need to be formed before accretion processes responsible of formation of chondrites. Of fundamental importance in these processes is the gradual and reversible liquid-glass transition (or glass transition) of chondrules from a soft and ductile state into a hard and brittle glassy state as the temperature decrease. Explanations based on post-accretionary processes, such as impacts or thermal metamorphism, are demonstrably inadequate for the observed ductile deformation of certain chondrules in ordinary chondrites (Metzler, 2012). Cooling of chondrules and chondrite accretion processes can thus be summarized from a mechanistic point of view for both ordinary and carbonaceous chondrites (Fig. 6). The key aspect is the evolution in mechanical properties quantified by the fact that melt/mesostasis dynamic viscosity increases upon cooling because of (i) the Arrhenian temperature dependence (ii) the accumulation of crystals (Lejeune and Richet, 1995) and (iii) the evolution of the residual melt composition toward a more polymerized state (Libourel et al., 2006). Too hot, partially molten chondrules will coalesce forming isolated objects, such as compound chondrules like two or more chondrules fused together (Weyrauch and Bischoff, 2012; Jacquet, 2021; Arakawa and Nakamoto, 2016), but not chondrites (Metzler, 2012). Based on our results, chondrites need to form at lower temperatures, above or below the glass transition temperature of chondrule mesostases. In the following we will discuss the differences in the mechanisms involved, and then the astrophysical/cosmochemical implications.

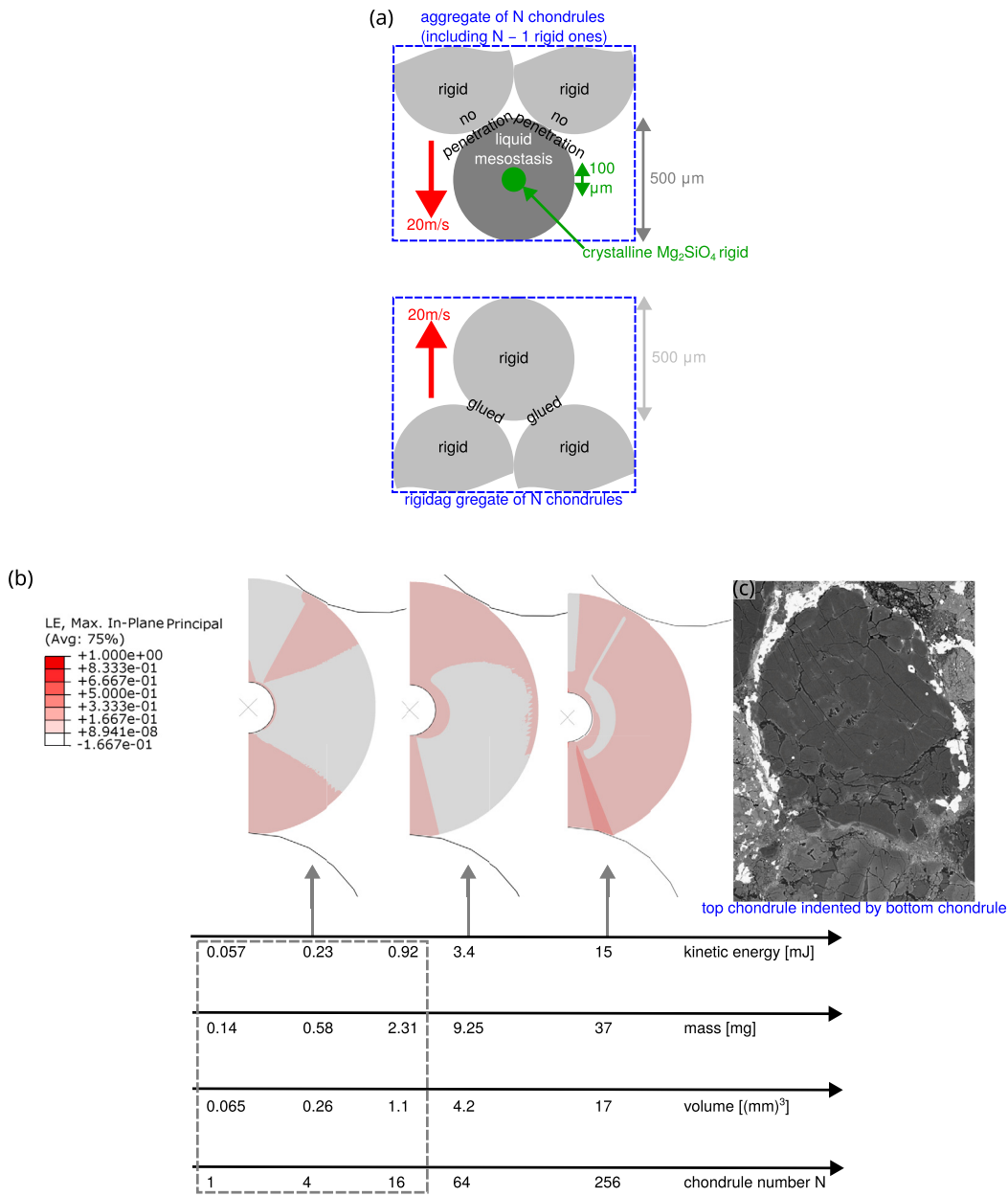

**Fig. 2.** (a) Collision model between two chondrule aggregates of identical number of chondrules used in the simulation at $40\,\mathrm{m\,s^{-1}}$ relative velocity. In the upper aggregate, the colliding chondrule presents a liquid/ductile mesostasis and other chondrules are fully rigid. In the bottom aggregate, all chondrules are fully rigid and glued together. (b) Resulting deformed shape and maximal eigenvalue of logarithmic strain (representing the accumulated strain for each point of matter along its path in the matter, and taking the maximal eigenvalue of the second-order tensor to display the maximal stretching strain over all possible viewing directions) of mesostasis of the central partially crystallized chondrule. (c) Scanning electron microscopy image of a deformed chondrule.

### 3.1. Hot and plastic deformation

Chondrules can be plastically deformable only in a temperature range of approximately 400 K above the $T_g$ of their mesostasis with viscosity roughly between $10^4$ to $10^{13}$ poises *i.e.* $1 \times 10^{12}$ Pa s corresponding to the working and annealing points of glassmakers, respectively (Richet et al., 2021). In this frame, our simulations showed that hot collisions (typically above $T_g + 400\,\mathrm{K}$) involving aggregates of several tens of chondrules (Fig. 2) instead of isolated chondrules are required to attain plastic deformations of chondrule mesostasis similar to those observed in ordinary chondrite (Metzler, 2012). Considering relative velocities generally associated with chondrules ($< 150\,\mathrm{m\,s^{-1}}$, (Ueda et al., 2001; Arakawa and Nakamoto, 2016, 2019; Arakawa et al., 2023)) our calculations show that isolated chondrules provide too low kinetic energy at impact. This kinetic energy first converts itself in internal energy as elastic strain energy, inducing density change, and then back in kinetic energy as a bouncing back of chondrules. As the chondrule aggregates become massive enough *e.g.*, $\sim 10^5$ chondrules equivalent to few grams (Fig. 3), they will carry sufficient kinetic energy to induce plastic deformation onto mesostases that are still above their glass transition temperature. Such velocities (Fig. 3) are still lower than values generally associated with chondrules ($< 150\,\mathrm{m\,s^{-1}}$ and references aforementioned). We therefore suggest as a testable hypothesis that the deformation of chondrules in a low kinetic energy regime would be controlled by the size of the chondrule aggregates. It is however important to mention that such results primarily depend on the material properties of the mesostasis, and notably its dynamic viscosity which itself depend (to the first order) on the temperature.

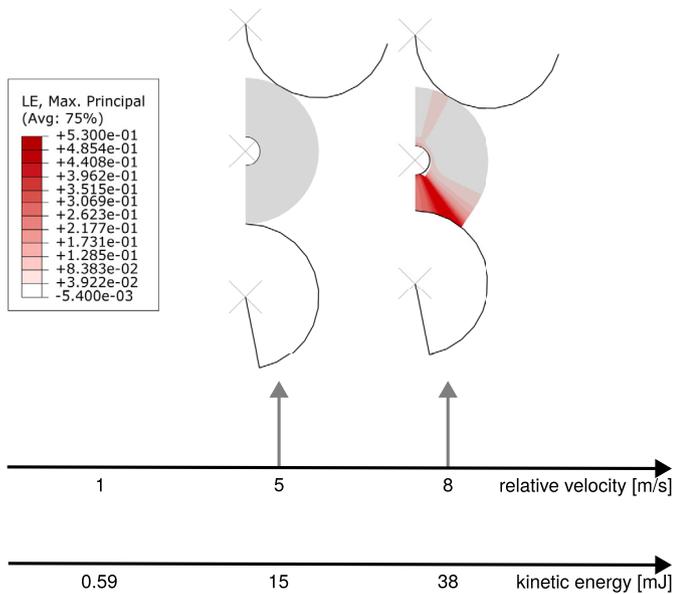

**Fig. 3.** Collision model between two chondrule aggregates of volume 1 cm³ and chondrule number $2^{14}$ as a function of the relative velocity. Resulting deformed shape and maximal eigenvalue of logarithmic strain (representing the accumulated strain for each point of matter along its path in the matter, and taking the maximal eigenvalue of the second-order tensor to display the maximal stretching strain over all possible viewing directions) of mesostasis of the central partially crystallized chondrule. For relative velocities of $1\,\mathrm{m\,s^{-1}}$ and $5\,\mathrm{m\,s^{-1}}$, the mesostasis of the chondrule is virtually not deformed so the $1\,\mathrm{m\,s^{-1}}$ case is not shown.

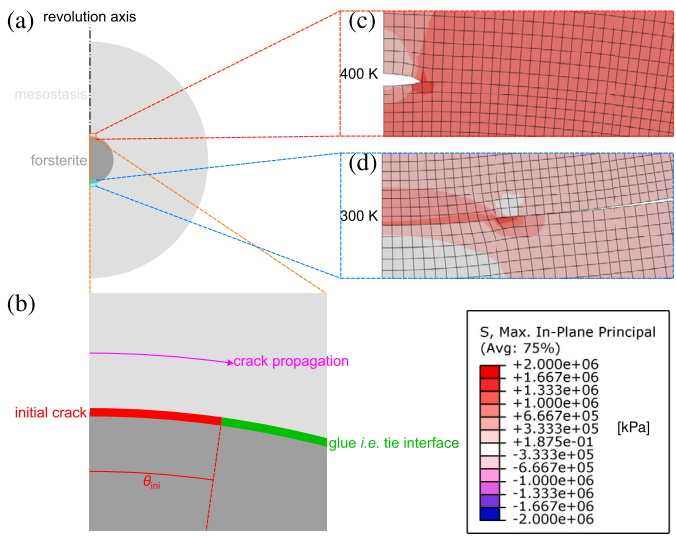

**Fig. 4.** (a) Model for crack propagation along the forsterite-mesostasis interface in the chondrule upon cooling; (b) zoom on the initial crack, where $\theta_{\mathrm{ini}}$ is the angular position measured from the vertical (orange lines) *i.e.* colatitude of the crack front in that initial position. (c, d). Maximal *i.e.* first principal stress field in the finite element model around the crack front for the standard-sized chondrule size for (c) the initial crack front position at 400 K and (d) penultimate crack front position at 300 K.

## 3.2. Cold cooling-dominated cracking

According to rock mechanics (Fredrich and Wong, 1986) and studies on volcanic rock fracturing (Browning et al., 2016), it is easier for cracks to form during cooling and contraction than during heating and expansion. This is because the thermal stresses generated by contraction are tensile stresses, which are more likely to cause cracking than compressive stresses. Thermal stresses are generated by two main mechanisms: mismatch in thermal expansion coefficients between different phases (see supplementary material and figure S3) and/or thermal expansion heterogeneity within single phases.

During cooling of chondrules, their heterogeneity of phases and thus of their thermal contraction induces stresses. Hence, the heterogeneous shrinking of chondrules when they cool below the glass transition temperature of their mesostasis is likely responsible for the cracks found in them.

Reinforcing this conclusion, quenched synthetic chondrules consistently exhibit cracks. For instance (Fig. 7), an olivine-rich synthetic chondrule (mass composition: 53 % SiO₂, 14 % Al₂O₃, 9 % CaO, 23 % MgO, 1 % Cr₂O₃; $T_g = 960\,\mathrm{K}$) heated to 1723 K for 2 h, cooled at $2\,\mathrm{K\,h^{-1}}$, and finally quenched in air at 1581 K displays cracks within and along the edges of olivine grains, some of which retain their euhedral morphology. This crack pattern is also observed in other quenched synthetic basalts (Mourey and Shea, 2019; Auxerre et al., 2022).

As olivines shrink more than their glassy mesostasis ($0.7\times10^{-5}\,\mathrm{K^{-1}} < \alpha_{\mathrm{olivine}} < 1.5\times10^{-5}\,\mathrm{K^{-1}}$ from supplementary figure S3, and $\alpha_{\mathrm{mesostasis}} \sim 4\times10^{-6}\,\mathrm{K^{-1}}$ from Methods and (Liu et al., 2020), where $\alpha_{\mathrm{olivine}}$ and $\alpha_{\mathrm{mesostasis}}$ are thermal expansion coefficients of olivine and mesostasis respectively), olivine-mesostasis interfaces undergo tensile stresses. Moreover, the edges and vertices of euhedral olivine-mesostasis interfaces can also create localized stresses and strain energy concentrations (Liu et al., 2015), which can promote further crack nucleation and propagation at olivine-mesostasis interfaces. As a result, chondrules typically have a largely isotropic crack orientation, with some cracks being guided by the olivine-mesostasis interface and others preserving the euhedral morphology of the chondrule olivine with some mesostasis attached (Fig. 7 and supplementary figure S6). The driving force of this autonomous and spontaneous fracturing process is the combination of (i) the amplitude of the temperature change and (ii) the inter-phase heterogeneity of shrinking upon cooling. Cracks can thus propagate by this mechanism irrespectively of the types (porphyritic or non-porphyritic), chemistry and cooling rate of multi-phased/crystallized chondrules.

In natural chondrule, the volume fraction of glassy mesostasis is usually smaller than one considered here, as the majority of the volume is occupied by olivine/ pyroxene crystal grains. To assess the influence of this characteristic, it is possible to consider that for a fixed chondrule diameter, increasing the olivine crystal ball diameter (and hence its volume fraction) should (i) *a priori* retain the same overall potential for storing internal elastic strain energy and thus for inducing/driving the crack propagation and (ii) increase the area of interphase (glassy mesostasis vs. olivine crystal) interface and thus internal interface energy to supply to break it/its atomic bonds. So overall crack propagation should be more difficult for larger olivine size at constant chondrule size.

The principles outlined above can be applied not only to olivines within chondrules but also to metal grains, exploring their potential role in spontaneous cracking. Consider pure iron and pure nickel, whose thermal expansion coefficients at 20 K ($11.8\times10^{-6}\,\mathrm{K^{-1}}$ and $13\times10^{-6}\,\mathrm{K^{-1}}$, respectively) exceed that of forsterite at the same temperature (figure S3c; (Yamaguchi et al., 2019)). Assuming this difference persists throughout the cooling process (1000 K to 300 K), the differential shrinkage between pure iron or nickel particles and the glassy mesostasis would be more pronounced. If these metal particles are smaller than olivine crystals crack propagation would be favored due to their size (see above). Consequently, the interface between iron/nickel particles and the glassy mesostasis would be more susceptible to crack propagation than the olivine-glassy mesostasis interface, owing to both size effects and differential thermal shrinkage. In a manner analogous to isolated olivines, this mechanism could explain the presence of isolated metal grains in the matrix, either with or without an attached layer of glassy ± olivine (Connolly et al., 2001; Jacquet et al., 2013; van Kooten et al., 2022).

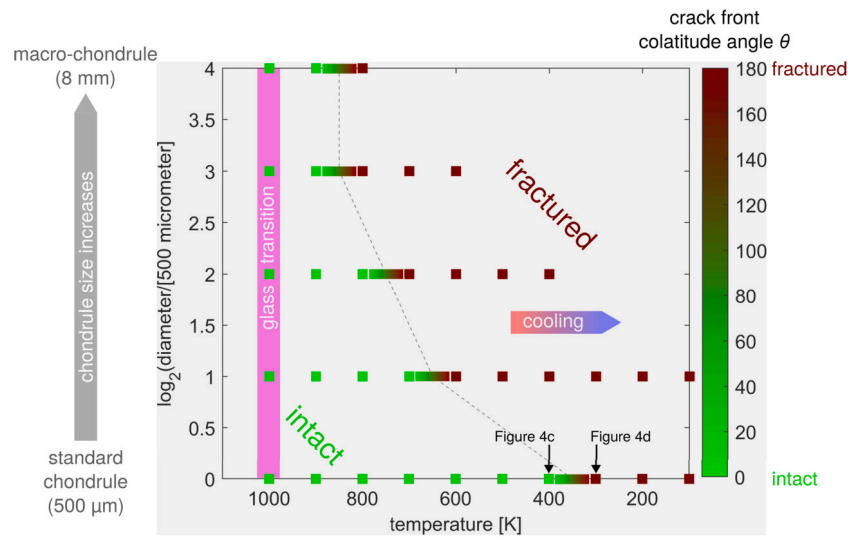

**Fig. 5.** Crack propagation upon cooling as a function of chondrule size, represented by the angular position *i.e.* colatitude $\theta$ of the crack front (color) as a function of temperature (abscissa) and chondrule size (ordinate, logarithmic scale to represent chondrule diameters of 500 μm, 1 mm, 2 mm, 4 mm and 8 mm) for the central forsterite. For each temperature, the dashed gray line connects the points at which half of the phase interface has cracked. The rectangle with a color gradient represents crack propagation within a temperature interval of at most 100 K.

### 3.3. Astrophysical and cosmochemical implications

In the case of ordinary chondrites, chondrule aggregates can only form in localized high-density regions of the disk (Cuzzi and Alexander, 2006; Alexander et al., 2008). These objects would then rapidly collapse and accrete into chondritic parent bodies, along with a very low abundance of fine-grained matrix (see Fig. 11 in (Metzler, 2012) showing that some chondrules are mantled by fine-grained dust; Siron et al., 2022). It is likely that one population of chondrules accreted at a high enough temperature to allow viscous deformation, while another population accreted after the temperature had dropped and the chondrules, and possibly some compound chondrules, had solidified into rigid spheres (Fig. 1). In this scenario, isolated chondrules first aggregate and grow by gentle collisions, forming snowflake-like aggregates. These aggregates become massive enough to carry sufficient kinetic energy to induce plastic deformation in the mesostases of chondrules that are still above their glass transition temperature (Fig. 6). Knowing that the amount of ductile deformation of a chondrule, measured in ordinary chondrites varies from 0 to 56 % (Metzler, 2012; Metzler and Pack, 2016; Ruzicka et al., 2020), it is tempting to relate this variability to the size of chondrule aggregates. The larger the aggregate, the higher the ductile deformation (Fig. 2). Undeformed or partly deformed chondrules that cool below $T_g$ are no longer ductile but now brittle and will instead fracture in response to subsequent thermal stresses (Fig. 6). However, fragmentation and comminution in ordinary chondrites seems to be limited by the abundance and the close-fit textures of chondrules as well as by the thermal energy released by the accretion of hot chondrules. This scenario of hot accretion argues in favor of the unity of place and time for the formation of chondrules in ordinary chondrites (Metzler, 2012). Following the same logic, the absence of deformed chondrules in carbonaceous chondrites could be attributed to the insufficient mass of chondrules per unit volume to form aggregates with enough kinetic energy for promoting ductile deformation.

As shown by our simulations, chondrule fragmentation could spontaneously occur in a tensile regime imposed by the thermal stress as the temperature drops below $T_g$. Even if chondrules are partially fractured (as in Fig. 7), very low-velocity collisions (*e.g.* much less than 8 m s$^{-1}$) fragment and destroy chondrules with pre-existing cracks and high glassy abundances (Ueda et al., 2001). Due to such efficiency, the liberated chondrule fragments must be dispersed in space and mixed with variable amounts of unprocessed particles and fine-grained CI dust, before the rapid collapse and accretion of these objects into chondritic parent bodies. The formation of carbonaceous chondrites from the accretion of colder materials than in the case of ordinary chondrites (Fig. 6) is likely a contributing factor to the higher modal abundance of chondrule fragments and notably isolated olivine grains in the former. This is because the larger amplitude of temperature change during cooling in carbonaceous chondrites, compared to ordinary chondrites (Ruzicka et al., 2024), leads to a more pronounced differential thermal contraction, which promotes chondrule cracking (Fig. 5). The higher porosity and larger voids/cracks observed recently in carbonaceous chondrite chondrules (Kadlag, 2023) support this finding. Whether such a larger amplitude of temperature change can be caused by a more efficient radiative cooling of chondrules in low-density regions of the disk is still to be determined (Dullemond et al., 2014; Delpeyrat et al., 2019).

This scenario of spontaneous cracking also has implications on (i) size-sorting of chondrules and isolated silicates and (ii) non-CI chemistry of the matrix in carbonaceous chondrites. It has long been recognized that chondrules in individual chondrites have populations with restricted size ranges (Krot et al., 2005). Hypotheses for the sorting process include chondrule sorting by mass and sorting by some aerodynamic mechanism (Dodd, 1975; Cuzzi et al., 2000; Metzler et al., 2019). Here it is proposed that spontaneous cold cracking during chondrule cooling must be also considered as an effective and ubiquitous sorting process. As chondrule size increases, the ratio of elastic strain energy to crack interface increases. This is because (assuming a constant chondrule size-to-olivine size ratio) the elastic strain energy is proportional to the volume and thus to the cube of the size due to differential thermal contraction, whereas the crack interface energy is proportional to the crack surface and thus to the square of the olivines size. This promotes cracking, which can lead to the natural sorting of chondrules by size through fracturing in different fragments (Fig. 5 and difference in chondrule size between CM or CO chondrites (Krot et al., 2005; Jones, 2012).

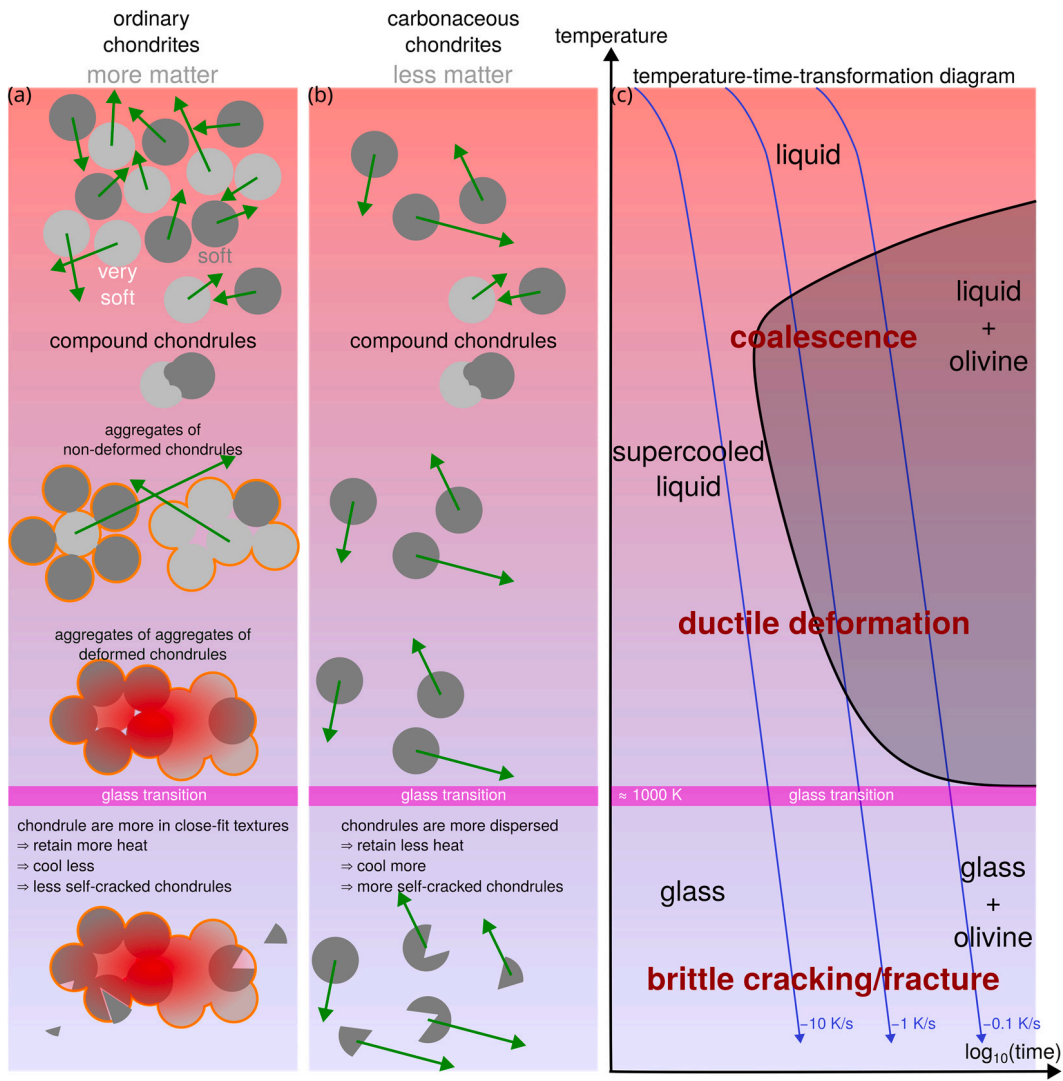

**Fig. 6.** Schematic (a, b) diagram and (c) time-temperature-transformation (TTT) illustrating chondrule cooling behavior for both (a) ordinary and (b) carbonaceous chondrites. In the TTT diagram, liquid domain, onset of crystallization, glass transition, glass and glass + olivine domains are indicated together with chondrule cooling rates (blue curves) as an indication (Desch et al., 2012; Jones et al., 2018). Glass transition temperature $T_g$ is estimated around 1000 K for chondrules. As chondrule temperature is decreased, there is a gradual and reversible liquid-glass transition from a soft and ductile state into a hard and brittle glassy state. Only the temperature difference matters in the self-cracking/fracturing process below $T_g$. Cracks can thus propagate in different types of chondrules (porphyritic, non-porphyritic, or glassy) independently of their cooling rates of formation. See text for further explanations.

Accordingly, larger chondrules are more likely to fracture than smaller chondrules, leaving large, isolated olivine grains as the only trace of these (formerly macro-) chondrules, some of which reaching 500 μm in diameter in carbonaceous chondrites (Fig. 1, supplementary figure S6, (Weinbruch et al., 2000; Pack et al., 2004)).

Cold cracking must also fragment chondrule mesostases and mix the resulting glass shards with the surrounding CI dust. This can cause the bulk chemistry of the resulting matrix to shift towards the specific composition of chondrule mesostases (CI-normalized Mg/Si/CI < 0.2, 2 < Al/Si/CI < 8 and 0.8 < Ca/Si/CI < 8 (Libourel et al., 2006)), with net Mg depletion and Al and Ca gain relative to CI as observed in carbonaceous chondrites (Patzer et al., 2021b,a, 2023). Such comminution process of the glassy mesostases may be the rationale of the non-CI like matrix in carbonaceous chondrites. Correction of this shift confirms the common origin and the cogenetic evolution of both chondrule and matrix components from a CI-like reservoir. However, this comminution effect must be more difficult to recognize in ordinary chondrites, in which the matrix is in lower abundance than in carbonaceous chondrites. It would be interesting to test whether comminuted glass shards could be related to amorphous silicate matter, agglutinate/gems-like material or any mixtures of alteration products found in most matrix of chondrites (Leroux et al., 2015; Dobrică and Brearley, 2020; Hewins et al., 2021).

## 4. Conclusion

In conclusion, our mechanistic approach provides insights into the history of the major chondritic components, chondrules and matrix during chondrite accretion. In particular, we show that hot ductile deformation and cold spontaneous cracking are universal processes that affect chondrules during cooling, regardless of the nature of the chondrule-forming heating event or the textural and chemical type of the chondrules. Consequently, ordinary and carbonaceous chondrites result from similar ductile and brittle low-energy processes, whose respective intensities depend on the amount of matter processed/number density of chondrules in the chondrite forming region and on the thermal environment *i.e.* hotter versus colder regions. In agreement with the view that chondrule-forming heating events and chondrite accretion were closely linked in time and space, this approach also suggests that chondrite accretion environments were thermally active and less turbulent than originally thought. It remains now to be understood how chondrule number density in excess to $1 \times 10^3$ m$^{-3}$ in high-density regions of the disk (Cuzzi and Alexander, 2006) align with accretion processes governed by low velocity/energy collision regimes as inferred by this study.

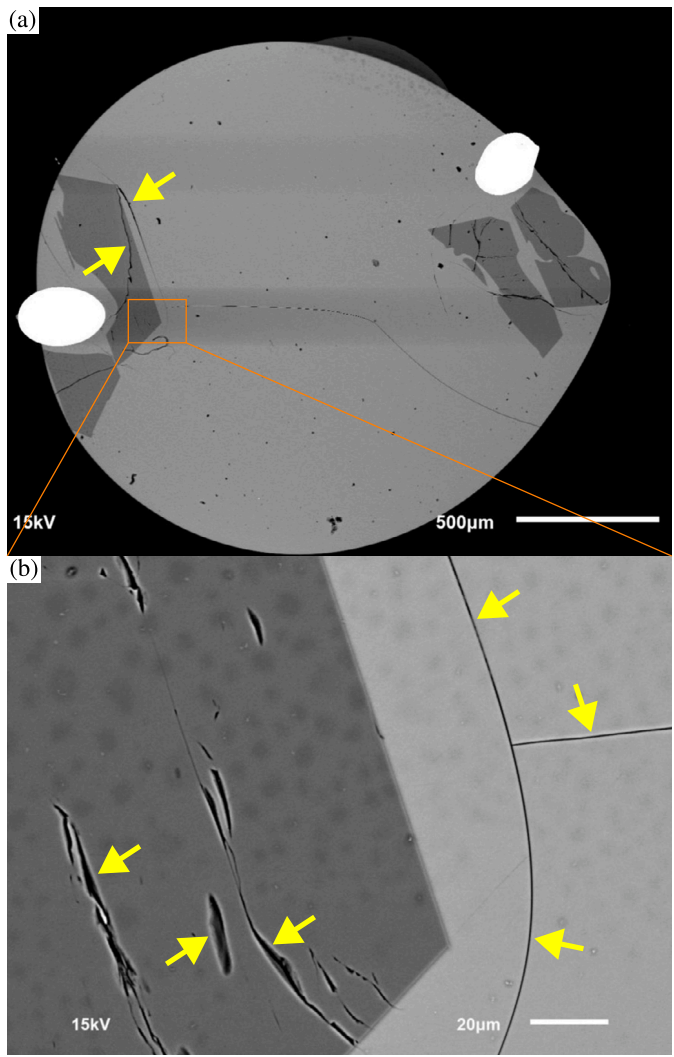

**Fig. 7.** Back scattered electron scanning electron microscope images of a quenched synthetic olivine-bearing chondrule showing cracks (yellow arrows) both in olivine and in the glassy mesostasis. (a) global view and (b) magnification at olivine-glass interface. The starting material (53 % $SiO_2$, 14 % $Al_2O_3$, 9 % CaO, 23 % MgO and 1 % $Cr_2O_3$, in weight fraction) have been heated to 1723 K for 2 h then cooled at 2 K h$^{-1}$ before a final quench in air at 1581 K. $T_g$ = 960 K is the glass transition temperature. See text for details.

## CRediT authorship contribution statement

**Anthony Seret:** Writing – review & editing, Writing – original draft, Visualization, Validation, Software, Methodology, Investigation, Formal analysis, Data curation, Conceptualization. **Guy Libourel:** Writing – review & editing, Writing – original draft, Visualization, Validation, Supervision, Resources, Project administration, Investigation, Funding acquisition, Formal analysis, Conceptualization.

## Funding

This project was financially supported by ANR O-Return ANR-21-CE49-0005 (GL and MP) and by ERC HolyEarth ERC Grant N 101019380.

## Declaration of competing interest

The authors declare that they have no known competing financial interests or personal relationships that could have appeared to influence the work reported in this paper.

## Acknowledgements

Morbidelli A., Michel P. (UniCA-OCA, Nice) are thanked for their advice and discussions at different levels of progress of this study. We are particularly grateful to Daniel Pino Munoz (CEMEF, Mines ParisTech, Nice) for his enlightened assistance in the conceptualization of these simulations. We also thank Laurent Tissandier (CRPG-CNRS, Nancy) for performing the experiment of chondrule quenching. Eventually, we thank Olivier Mousis for his editorial handling of the manuscript, and Elishevah van Kooten for an anonymous reviewer for their reviews.

## Appendix A. Supplementary material

Supplementary material related to this article can be found online at https://doi.org/10.1016/j.epsl.2024.119066.

## Data availability

Data will be made available on request.

## References

Alexander, C.M.O., Grossman, J.N., Ebel, D.S., Ciesla, F.J., 2008. The formation conditions of chondrules and chondrites. Science 320 (5883), 1617–1619. https://www.science.org/doi/abs/10.1126/science.1156561.

Arakawa, S., Nakamoto, T., 2016. Compound chondrule formation via collision of supercooled droplets. Icarus 276, 102–106. https://www.sciencedirect.com/science/article/pii/S0019103516301245.

Arakawa, S., Nakamoto, T., 2019. Compound chondrule formation in optically thin shock waves. Astrophys. J. 877 (2), 84–99. https://doi.org/10.3847/1538-4357/ab1b3e.

Arakawa, S., Yamamoto, D., Ushikubo, T., Kaneko, H., Tanaka, H., Hirose, S., Nakamoto, T., 2023. Oxygen isotope exchange between molten silicate spherules and ambient water vapor with nonzero relative velocity: implication for chondrule formation environment. Icarus 405, 115690. https://www.sciencedirect.com/science/article/pii/S0019103523002671.

Auxerre, M., Faure, F., Lequin, D., 2022. The effects of superheating and cooling rate on olivine growth in chondritic liquid. Meteorit. Planet. Sci. 57, 1474–1495.

Browning, J., Meredith, P., Gudmundsson, A., 2016. Cooling dominated cracking in thermally stressed volcanic rocks. Geophys. Res. Lett. 43, 8417–8425.

Connelly, J.N., Bizzarro, M., Krot, A.N., Nordlund, Å., Wielandt, D., Ivanova, M.A., 2012. The absolute chronology and thermal processing of solids in the solar protoplanetary disk. Science 338 (6107), 651–655.

Connelly, H.C., Huss, G.R., Wasserburg, G.J., 2001. On the formation of ferni metal in renazzo-like carbonaceous chondrites. Geochim. Cosmochim. Acta 65 (24), 4567–4588. https://www.sciencedirect.com/science/article/pii/S0016703701007499.

Cuzzi, J., Alexander, C., 2006. Chondrule formation in particle-rich nebular regions at least hundreds of kilometers across. Nature 441, 483–485.

Cuzzi, J., Hogan, R., Paque, J., Dobrovolskis, A., 2000. Size-selective concentration of chondrules and other small particles in protoplanetary nebula turbulence. Astrophys. J. 546, 496–508.

Delpeyrat, J., Pigeonneau, F., Libourel, G., 2019. Chondrule radiative cooling in a non-uniform density environment. Icarus 329, 1–7. https://www.sciencedirect.com/science/article/pii/S0019103518305505.

Desch, S., Morris, M., Connolly, H., Boss, A., 2012. The importance of experiments: constraints on chondrule formation models. Meteorit. Planet. Sci. 47, 1139–1156.

Dobrică, E., Brearley, A., 2020. Amorphous silicates in the matrix of semarkona: the first evidence for the localized preservation of pristine matrix materials in the most unequilibrated ordinary chondrites. Meteorit. Planet. Sci. 55, 649–668.

Dodd, R.T., 1975. Accretion of the ordinary chondrites. Earth Planet. Sci. Lett. 30, 281–291. https://api.semanticscholar.org/CorpusID:129842218.

Doyle, P., Jogo, K., Nagashima, K., Krot, A., Wakita, S., Ciesla, F., Hutcheon, I., 2015. Early aqueous activity on the ordinary and carbonaceous chondrite parent bodies recorded by fayalite. Nat. Commun. 6, 7444.

Dullemond, C.P., Stammler, S.M., Johansen, A., 2014. Forming chondrules in impact splashes. i. radiative cooling model. Astrophys. J. 794 (1), 91–102. https://doi.org/10.1088/0004-637X/794/1/91.

Fredrich, J.T., Wong, T., 1986. Micromechanics of thermally induced cracking in three crustal rocks. J. Geophys. Res. 91, 12743–12764. https://api.semanticscholar.org/CorpusID:128646045.

Goudy, S.P., 2019. Assessment of cluster chondrite accretion temperature using electron backscatter diffraction and implications for chondrule formation models. Master's thesis. Portland State University.

Griffith, A.A., 1920. The phenomena of rupture and flow in solids. Philos. Trans. R. Soc. Lond. A 221, 163–198.

Hewins, R., Jr, C., Libourel, G., 2005. Experimental constraints on chondrule formation. In: Chondrites and the Protoplanetary Disk, vol. 341, p. 286.

Hewins, R., Zanetta, P.-M., Zanda, B., Guillou, C., Gattacceca, J., Sognzoni, C., Pont, S., Piani, L., Rigaudier, T., Leroux, H., Brunetto, R., Maupin, R., Djouadi, Z., Bernard, S., Deldicque, D., Malarewicz, V., Dionnet, Z., Aleon-Toppani, A., King, A., Borondics, F., 2021. Northwest Africa (nwa) 12563 and ungrouped c2 chondrites: alteration styles and relationships to asteroids. Geochim. Cosmochim. Acta 311, 238–273.

Holmén, B.A., Wood, J.A., 1986. Chondrules that indent one another: evidence for hot accretion? Meteoritics 21.

Jacquet, E., 2021. Collisions and compositional variability in chondrule-forming events. Geochim. Cosmochim. Acta 296, 18–37. https://www.sciencedirect.com/science/article/pii/S0016703720307389.

Jacquet, E., Paulhiac-Pison, M., Alard, O., Kearsley, A.T., Gounelle, M., 2013. Trace element geochemistry of cr chondrite metal. Meteorit. Planet. Sci. 48 (10), 1981–1999. https://onlinelibrary.wiley.com/doi/abs/10.1111/maps.12212.

Jones, R.H., 2012. Petrographic constraints on the diversity of chondrule reservoirs in the protoplanetary disk. Meteorit. Planet. Sci. 47 (7), 1176–1190. https://onlinelibrary.wiley.com/doi/abs/10.1111/j.1945-5100.2011.01327.x.

Jones, R.H., Villeneuve, J., Libourel, G., 2018. Chondrules: Records of Protoplanetary Disk Processes. Cambridge University Press, pp. 57–90. Ch. Thermal Histories of Chondrules.

Kadlag, Y., 2023. Physical properties and average atomic numbers of chondrules using computed tomography. Planet. Space Sci. 238, 3–4.

Krot, A.N., Scott, E.R.D., Reipurth, B. (Eds.), 2005. Chondrites and the Protoplanetary Disk. Astronomical Society of Pacific Conference Series, vol. 341. 390 Ashton Avenue San Francisco, California 94112.

Kruijer, T., Kleine, T., Borg, L., 2020. The great isotopic dichotomy of the early solar system. Nat. Astron. 4, 32–40.

Lejeune, A.-M., Richet, P., 1995. Rheology of crystal-bearing silicate melts: an experimental study at high viscosities. J. Geophys. Res. 100, 4215–4229. https://api.semanticscholar.org/CorpusID:130684987.

Leroux, H., Cuvillier, P., Zanda, B., Hewins, R., 2015. Gems-like material in the matrix of the Paris meteorite and the early stages of alteration of cm chondrites. Geochim. Cosmochim. Acta 170, 247–265.

Libourel, G., Krot, A.N., Tissandier, L., 2006. Role of gas-melt interaction during chondrule formation. Earth Planet. Sci. Lett. 251 (3), 232–240. https://www.sciencedirect.com/science/article/pii/S0012821X06006455.

Liu, J., Chang, Z., Wang, L., Xu, J., Kuang, R., Wu, Z., 2020. Exploration of basalt glasses as high-temperature sensible heat storage materials. ACS Omega 5, 19236–19246.

Liu, M., Gan, Y., Hanaor, D.A., Liu, B., Chen, C., 2015. An improved semi-analytical solution for stress at round-tip notches. Eng. Fract. Mech. 149, 134–143. https://www.sciencedirect.com/science/article/pii/S0013794415005834.

Lodders, K., 2003. Solar system abundances and condensation temperatures of the elements. Astrophys. J. 591 (2), 1220–1247. https://doi.org/10.1086/375492.

Metzler, K., 2012. Ultrarapid chondrite formation by hot chondrule accretion? Evidence from unequilibrated ordinary chondrites. Meteorit. Planet. Sci. 47 (12), 2193–2217. https://onlinelibrary.wiley.com/doi/abs/10.1111/maps.12009.

Metzler, K., Hezel, D., Nellessen, J., 2019. Various size-sorting processes for millimeter-sized particles in the sun's protoplanetary disk? Evidence from chondrules in ordinary chondrites. Astrophys. J. 887, 230–239.

Metzler, K., Pack, A., 2016. Chemistry and oxygen isotopic composition of cluster chondrite clasts and their components in ll3 chondrites. Meteorit. Planet. Sci. 51, 276–302.

Mourey, A., Shea, T., 2019. Forming olivine phenocrysts in basalt: a 3d characterization of growth rates in laboratory experiments. Front. Earth Sci. 7.

Nelson, V.E., Rubin, A.E., 2010. Size-frequency distributions of chondrules and chondrule fragments in ll3 chondrites: implications for parent-body fragmentation of chondrules. Meteorit. Planet. Sci. 37 (10), 1361–1376. https://onlinelibrary.wiley.com/doi/abs/10.1111/j.1945-5100.2002.tb01034.x.

Pack, A., Yurimoto, H., Palme, H., 2004. Petrographic and oxygen-isotopic study of refractory forsterites from r-chondrite dar al gani 013 (r3.5-6), unequilibrated ordinary and carbonaceous chondrites. Geochim. Cosmochim. Acta 68, 1135–1157.

Patzer, A., Bullock, E., Alexander, C., 2021a. Testing models for the compositions of chondrites and their components: ii. cr chondrites. Geochim. Cosmochim. Acta 319, 1–29.

Patzer, A., Bullock, E., Alexander, C., 2023. Testing models for the compositions of chondrites and their components: iii. cm chondrites. Geochim. Cosmochim. Acta 359, 30–45.

Patzer, A., Bullock, E.S., Alexander, C.M.O., 2021b. Testing models for the compositions of chondrites and their components: i. co chondrites. Geochim. Cosmochim. Acta 304, 119–140. https://www.sciencedirect.com/science/article/pii/S0016703721002192.

Richet, P., Conradt, R., Takada, A., Dyon, J. (Eds.), 2021. Encyclopedia of Glass Science, Technology, History, and Culture. The American Ceramic Society, 550 Polaris Pkwy, Ste 510, Westerville, OH 43082.

Ruzicka, A., Hugo, R., Hutson, M., 2015. Deformation and thermal histories of ordinary chondrites: evidence for post-deformation annealing and syn-metamorphic shock. Geochim. Cosmochim. Acta 163, 219–233. https://www.sciencedirect.com/science/article/pii/S0016703715002367.

Ruzicka, A.M., Goudy, S.P., Hugo, R.C., 2020. Role of hot accretion and deformation in producing cluster and type 3 ordinary chondrites. In: Proceedings of the 51th Lunar and Planetary Science Conference, p. 1308. https://api.semanticscholar.org/CorpusID:211142997.

Ruzicka, A.M., Hugo, R.C., Friedrich, J.M., Ream, M.T., 2024. Accretion of warm chondrules in weakly metamorphosed ordinary chondrites and their subsequent reprocessing. Geochim. Cosmochim. Acta 378, 1–35. https://www.sciencedirect.com/science/article/pii/S0016703724002631.

Scott, E.R.D., Krot, A.N., 2014. 1.2 - chondrites and their components. In: Holland, H.D., Turekian, K.K. (Eds.), Treatise on Geochemistry, second edition. Elsevier, Oxford, pp. 65–137. https://www.sciencedirect.com/science/article/pii/B9780080959757001042.

Siron, G., Fukuda, K., Kimura, M., Kita, N.T., 2022. High precision 26al-26mg chronology of chondrules in unequilibrated ordinary chondrites: evidence for restricted formation ages. Geochim. Cosmochim. Acta 324, 312–345. https://www.sciencedirect.com/science/article/pii/S0016703722000758.

Ueda, T., Murakami, Y., Ishitsu, N., Kawabe, H., Inoue, R., Nakamura, T., Sekiya, M., Takaoka, N., 2001. Collisional destruction experiment of chondrules and formation of fragments in the solar nebula. Earth Planets Space 53, 927–935. https://doi.org/10.1186/BF03351689.

van Kooten, E.M., Kubik, E., Siebert, J., Heredia, B.D., Thomsen, T.B., Moynier, F., 2022. Metal compositions of carbonaceous chondrites. Geochim. Cosmochim. Acta 321, 52–77. https://www.sciencedirect.com/science/article/pii/S0016703722000175.

Villeneuve, J., Chaussidon, M., Libourel, G., 2009. Homogeneous distribution of 26al in the solar system from the mg isotopic composition of chondrules. Science 325 (5943), 985–988.

Warren, P.H., 2011. Stable-isotopic anomalies and the accretionary assemblage of the Earth and Mars: a subordinate role for carbonaceous chondrites. Earth Planet. Sci. Lett. 311 (1), 93–100. https://www.sciencedirect.com/science/article/pii/S0012821X11005115.

Weinbruch, S., Palme, H., Spettel, B., 2000. Refractory forsterite in primitive meteorites: condensates from the solar nebula? Meteorit. Planet. Sci. 35, 161–171.

Weyrauch, M., Bischoff, A., 2012. Macrochondrules in chondrites—formation by melting of mega-sized dust aggregates and/or by rapid collisions at high temperatures? Meteorit. Planet. Sci. 47, 2237–2250.

Yamaguchi, A., Kimura, M., Barrat, J.-A., Greenwood, R., 2019. Compositional diversity of ordinary chondrites inferred from petrology, bulk chemical, and oxygen isotopic compositions of the lowest feo ordinary chondrite, yamato 982717. Meteorit. Planet. Sci. 54, 1919–1929.